\documentclass[conference]{IEEEtran}
\IEEEoverridecommandlockouts
\usepackage{graphicx}
\usepackage{amsthm}
\usepackage{bm}
\usepackage{color}
\usepackage{epsfig}
\usepackage{amsmath}
\usepackage{amssymb}
\usepackage{cite}
\usepackage{amsmath,amssymb,amsfonts}
\usepackage{algorithmic}
\usepackage{xcolor}
\usepackage{gensymb}
\usepackage{epstopdf}
\DeclareMathOperator{\atantwo}{atan_2}

\usepackage{fancyhdr}
\begin{document}
\bstctlcite{IEEEexample:BSTcontrol}
\title{Detecting Orbital Angular Momentum of Light in Satellite-to-Ground Quantum Communications \\
\vspace{-10pt}}
\author{\IEEEauthorblockN{Ziqing Wang$^1$, Robert Malaney$^1$, and Jonathan Green$^{2}$\\}
	\IEEEauthorblockA{
		\textit{$^1$School of Electrical Engineering \& Telecommunications, The University of New South Wales, Sydney, NSW 2052, Australia} \\
		\textit{$^{2}$Northrop Grumman Mission Systems, San Diego, California, USA}\\
		$^{1}$ziqing.wang1@student.unsw.edu.au, r.malaney@unsw.edu.au $^{2}$Jonathan.Green@ngc.com\\
	}
}

\maketitle
\thispagestyle{fancy}

\begin{abstract}
Satellite-based quantum communications enable a bright future for global-scale information security. However, the spin angular momentum of light, currently used in many mainstream quantum communication systems, only allows for quantum encoding in a two-dimensional Hilbert space. The orbital angular momentum (OAM) of light, on the other hand, enables quantum encoding in higher-dimensional Hilbert spaces, opening up new opportunities for high-capacity quantum communications. Due to its turbulence-induced decoherence effects, however, the atmospheric channel may limit the practical usage of OAM. In order to determine whether OAM is useful for satellite-based quantum communications, we numerically investigate the detection likelihoods for OAM states that traverse satellite-to-ground channels. We show that the use of OAM through such channels is in fact feasible. We use our new results to then investigate design specifications that could improve OAM detection---particularly the use of advanced adaptive optics techniques. Finally, we discuss how our work provides new insights into future implementations of space-based OAM systems within the context of quantum communications.
\end{abstract}
\section{Introduction}
Quantum communications can provide security at a level beyond that achievable via classical-only communications~\cite{Neda_Survey}. As one of the most important applications in quantum communications, Quantum Key Distribution (QKD) has been proved to provide unconditional security~\cite{unconditional_security}. In recent years, satellite-based quantum-information protocols have been intensively studied in order to extend the achievable communication range between terrestrial stations~\cite{Sat_QKD_research_2002,Sat_QKD_research_1,Sat_QKD_research_2,Sat_QKD_research_3,Sat_QKD_research_4,Sat_QKD_research_5}. Such studies have brought forth exciting experiments that have fully demonstrated the feasibility of satellite-based long-range quantum communications \cite{China1200km,proj_Micius}. Furthermore, recent demonstrations using low-cost micro-satellites~\cite{micro_satellite_50kg,nano_satellite} have illustrated a possible path towards global-scale quantum communications%

As a topological quantum number that is conserved during propagation, the orbital angular momentum (OAM) of light can take any integer value~\cite{Allen92}. Correspondingly, the OAM eigenstates form an infinite set of  orthonormal basis~\cite{OAM_Survey1}. Unlike the spin angular momentum (SAM) of light (which has been intensively used in Discrete-Variable (DV) quantum systems) that allows for quantum encoding in a two-dimensional Hilbert space, the OAM of light enables information encoding in a higher (theoretically, infinite) dimensional Hilbert space, opening up new opportunities for high-capacity quantum communications\cite{OAM_MUB1,OAM_MUB2,OAM_Bell_superdense}. It is well known that the OAM of light is conserved within the Spontaneous Parametric Down Conversion (SPDC) process, and thus can be entangled~\cite{OAM_entanglement,OAM_entanglement1,OAM_entanglement2}. According to a recent experiment, it could be argued that OAM entanglement distribution could be feasible over an FSO channel of more than 100km~\cite{OAM_143km}. The feasibility of utilizing OAM in existing QKD protocols has been both theoretically studied~\cite{OAM_MUB1,OAM_MUB2,OAM_Bell_superdense}, and experimentally demonstrated over a short Free Space Optical (FSO) channel of 210m~\cite{OAM_QKD_FSO_short}, and in a short fiber channel of 1.2km~\cite{QAM_QKD_6state_fibre}. Also, the distribution of OAM-based entanglement has been demonstrated over a turbulent FSO channel of 3km~\cite{OAM_entanglement_3km}.  However, existing theoretical and experimental research on OAM-based quantum communications have not paid sufficient attention to the realistic context of the satellite-based deployment. As such, the feasibility of OAM in such a context is still not clear.%

The atmospheric turbulence leads to refractive index fluctuations, imposing undesirable effects on an optical beam as it propagates within the atmospheric channel. These effects can cause crosstalk between OAM eigenstates, degrade the orthogonality among received OAM states, and eventually reduce the OAM detection performance~\cite{PatersonSPS}. The transmission of OAM-beams through atmospheric channels has been intensively investigated in the context of terrestrial FSO channels, e.g.~\cite{PatersonSPS,Gopaul07,Tyler09,Rodenburg12,Malik12,Alonso13,Ren_13,Rodenburg14,OptProbDet_18}. However, most of the previous works ignored the atmospheric turbulence effect on the OAM-beam's intensity profile. The atmospheric turbulence effect on the non-OAM-beam's intensity profile was modeled in~\cite{beam_wandering2016} in order to calculate the probability distribution of transmittance (PDT), however, the atmospheric turbulence effect on the non-OAM-beam's phase profile was ignored during the analysis, since the calculation of PDT did not require any information on the phase profile.

The propagation of OAM-beams traversing realistic satellite-to-ground atmospheric channels has not been studied so far. In this work, we remedy this situation and determine the OAM detection performance in satellite-to-ground quantum communications. The main contributions of this work are listed as follows. (i) First, we determine a realistic model for the satellite-to-ground atmospheric channel. (ii) Then, we apply this model to OAM-beams by considering the atmospheric turbulence effect on both their intensity profiles and their phase profiles. (iii) Finally, we numerically determine the OAM detection performance using Monte Carlo simulations. Our most important finding is that OAM is indeed feasible in satellite-to-ground quantum communications. This is largely due to the fact that the turbulence-induced degradation in the OAM detection performance can be alleviated by placing the ground station at a higher altitude, choosing a suitable wavelength, and\slash or adopting advanced real-time AO techniques.%

The structure of the remainder of this paper is as follows. In Section~\ref{ch:Sys}, we describe our system model of a satellite-to-ground quantum communication system utilizing OAM for quantum encoding. In Section~\ref{channel_model_top}, we describe the realistic atmospheric channel between the satellite and the ground station. In Section~\ref{ch:Sim} we describe our simulation settings, and present our simulation results. In Section~V we draw our conclusions.
\section{System Model} \label{ch:Sys}%
\subsection{System settings}
\begin{figure}[ht!]
	\centering
	\vspace{-13pt}
	\includegraphics[scale=0.62]{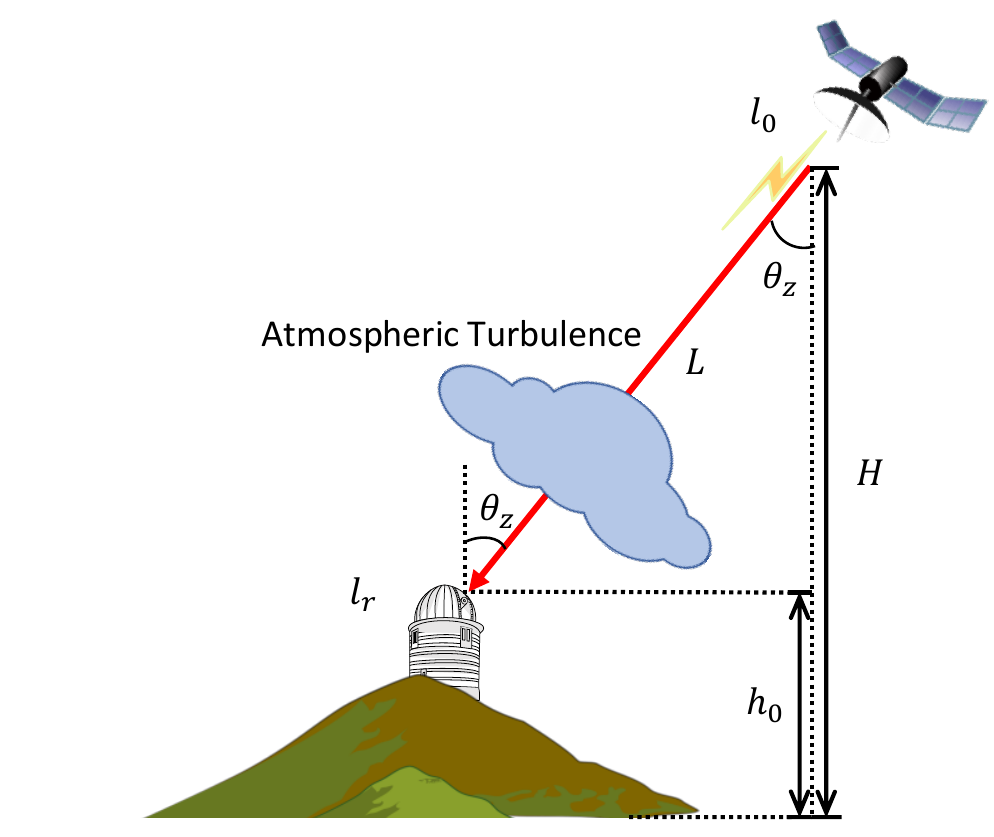}
	\vspace{-10pt}
	\caption{The system settings.}
	\vspace{-11pt}
	\label{fig:sysmodel}
\end{figure}

In this work, we consider the scenario of satellite-to-ground quantum communications which is illustrated in Fig.~(\ref{fig:sysmodel}). Here we briefly introduce our system settings. A satellite (at an altitude of $H$ and a zenith angle of $\theta_z$) sends an OAM eigenstate with the original OAM quantum number $l_0$ to the ground station (whose altitude is $h_0$) through an atmospheric channel (with a channel distance of $L$). For a higher-dimensional quantum encoding, $l_0$ can be chosen from a Hilbert space, $\{-l_{\rm{max}}, \ldots, +l_{\rm{max}}\}$, whose dimension is $(2l_{\rm{max}}+1)$. The ground station performs a quantum measurement on the received state and obtains a result $l_r$. Throughout this paper, we assume $\theta_z<45 \degree$ as discussed in~\cite{book} in order to satisfy the condition that the atmospheric turbulence within the satellite-to-ground atmospheric channel is weak. We denote the aperture radius at the transmitter and the receiver as $r_{t}$ and $r_{a}$, respectively.

\subsection{OAM eigenstates}
In cylindrical coordinates, the general form of an OAM eigenstate\footnote{Note, in this work our OAM-beams are in fact classical solutions to Maxwell's equations, and in this sense `classical' states. Even though we will use some formalism from quantum mechanics in our calculations, none of the states we investigate  will have negative Wigner functions (e.g. entangled states), and therefore are not formally `non-classical' states. However,  it can be shown that these  solutions for OAM-beams map to single photons carrying $l\hbar$ of OAM and therefore form an eigenbasis for single-photon descriptions. } is given by
\begin{equation}\label{Eq.OAM_Eigenstate}
\varphi _ { p , l } ( r , \theta,z) = R _ {p,l} ( r,z ) \frac{ \exp ( i l \theta ) }{\sqrt{2\pi}},
\end{equation}
where $r$ and $\theta$ are the radial and azimuthal coordinates, respectively, $z$ is the longitudinal coordinate, $l$ is the OAM quantum number, $p$ is the radial node number, and $ R_{p,l} (r,z)$ is the radial profile. In general, OAM eigenstates with different $l$ values are mutually orthogonal. In this paper, we choose $ R_{p,l} (r,z)$ to be Laguerre-Gauss functions, making OAM eigenstates correspond to the Laguerre-Gaussian ($LG$) mode set~\cite{Allen92}. $R_{p,l} (r,z)$  can be expressed by
\begin{equation}\label{LG}
\begin{aligned} R_{p,l}(r,z)= & 2 \sqrt{\frac{p !}{(p+|l|) !}} \frac { 1 } { w ( z ) } \left[ \frac { r \sqrt { 2 } } { w ( z ) } \right] ^ { | l | } \exp \left[ \frac { - r ^ { 2 } } { w ^ { 2 } ( z ) } \right] \\
& L _ { p } ^ { |l|} \left( \frac { 2 r ^ { 2 } } { w ^ { 2 } ( z ) } \right) \exp \left[ \frac { i k r ^ { 2 } z } { 2 \left( z ^ { 2 } + z _ { R } ^ { 2 } \right) } \right] \\
&\exp \left[ - i ( 2 p + | l | + 1 ) \tan ^ { - 1 } \left( \frac { z } { z _ { R } } \right) \right],
\end{aligned}
\end{equation}
where $w(z)=w_{0} \sqrt{1+(z / z_{R})^{2}}$ with $w_0$ being the beam-waist radius, $z_{\mathrm{R}}=\pi w_{0}^{2}/\lambda$ is the Rayleigh range, $\lambda$ is the optical wavelength,  $k=2\pi/\lambda$ is the optical wavenumber, and $L _ { p } ^ { | l | } ( x )$ is the generalized Laguerre polynomial~\cite{OAM_Survey1}. In this work, we denote the $LG$ modes and their complex amplitudes as ${LG}_{pl}$, and we set $p=0$ for the transmitted OAM eigenstates, since OAM is only related to $l$.

In general, the power of an $LG_{0l}$ mode is distributed within an infinite transverse plane. However, in practice, we always have a finite aperture size at any transceiver. $LG_{0l}$ modes are single-ring annular modes whose maximum optical intensity is located at $r=\sqrt{|l| / 2} w(z)$~\cite{OAM_Divergence}, indicating an $l$-dependent beam size. In this work, we use a finite quantity, $r_{\textit{0l}}(z)$, within which most ($\sim 90\%$) of the optical power is distributed, to characterize the beam size of the $LG_{0l}$ mode. $r_{\textit{0l}}(z)$ is given by
\begin{equation}\label{r_LG}
r_{\textit{0l}}(z)=\sqrt{2}r_{\rm{rms}}^\textit{{0l}}(z),
\end{equation}
where $r_{\rm{rms}}^\textit{{0l}}(z)$ is the rms beam radius, which can be evaluated by~\cite{OAM_rms}
\begin{equation}\label{r_rms}
\begin{aligned}
r_{\mathrm{rms}}^\textit{{0l}}(z)&= \sqrt{\int \!\!\!\!\int \! \varphi _ { 0 , l } ( r,\theta,z)\varphi_{ 0 , l }^{*} ( r,\theta,z) r \, \mathrm{d} r \mathrm{d} \theta} \\ &=\sqrt{\frac{|l|+1}{2}} w(z).
\end{aligned}
\end{equation}
In practice, the actual transmitter aperture radius $r_{t}$, and the actual receiver aperture radius $r_{a}$ should be set at least equal to $r_{0l}(0)$, and $r_{0l}(z)$, respectively.

\subsection{OAM detection probability}
In our system model, the transmitter aperture and the receiver aperture are situated at the points $z=0$ and $z=L$, respectively. The transmitted OAM eigenstate will be perturbed by the atmospheric turbulence while propagating. We denote the perturbed state at the ground station as $\Psi(r, \theta, z)$, which is generally a superposition of OAM eigenstates~\cite{PatersonSPS}, thus is not orthogonal to any OAM eigenstate. Based on Born's law, we express the conditional probability of obtaining a measurement $l_r$ as
\begin{equation}\label{ConProb}
P ( l_r | \Psi ) = \sum_{p}\left|a_{p, l_r}(z)\right|^{2}.
\end{equation}
where $a _ {p, l_r } ( z )$ is the probability amplitude which is given by
\begin{equation}\label{ProjectionCoeff}
a_{p,l_r}(z)=\left\langle\varphi_{p, l_r} | \Psi(r, \theta, z)\right\rangle.
\end{equation}
The rotational field correlation (RFC) function of the received OAM-beam can be defined as~\cite{PatersonSPS}
\begin{equation}\label{Eq.RFC}
C _ { \Psi } ( r , \theta , \theta', z ) = \left\langle \Psi ^ { * } ( r , \theta', z ) \Psi ( r , \theta , z ) \right\rangle,
\end{equation}
where $\theta$ and $\theta'$ denote two arbitrary azimuthal coordinates at $r$ in the transverse plane, and $\langle ... \rangle$ denotes an ensemble average. The unconditional probability of detecting $l_r$ can now be given by~\cite{PatersonSPS}
\begin{equation}\label{Eq.ProbFinal}
\begin{aligned}
P(l_r)=&\left\langle P ( l_r | \Psi ) \right\rangle \\
=&\int \!\!\!\! \int \!\! \int \! C _ { \Psi } ( r , \theta , \theta', z ) r d r \frac{\exp \left[-i l_r\left(\theta-\theta^{\prime}\right)\right]}{2 \pi} d \theta^{\prime} d \theta.
\end{aligned}
\end{equation}
The probability $P(l_0)$ gives the probability of correctly detecting $l_0$, and the probability $P(l_r)$ gives the probability of detecting $l_r$ $(l_r \ne l_0)$. In this work, when we refer to the `detection probability', we will be referring to the value $P(l_0)$. When we refer to the `crosstalk' between OAM eigenstates, we shall be referring to the value $P(l_r)$.
\section{Satellite-to-ground Atmospheric Channels}\label{channel_model_top}
\subsection{Atmospheric turbulence effect}\label{channel_model}
The effect of atmospheric turbulence on an optical beam manifests itself in both the beam's intensity profile and the beam's phase profile. The strength of the refractive index fluctuations within a satellite-based atmospheric channel can be characterized by the structure parameter $C_n^2(h)$. $C_n^2(h)$ can be described by the widely used Hufnagel-Valley (HV) model~\cite{HV}
\begin{equation}\label{Eq.HV}
\begin{aligned}
C_{n}^{2}(h)& = 0.00594(v_{\rm{rms}}/27)^{2}\left(h \times 10^{-5}\right)^{10} \exp{(-h/1000)} \\
& \ \ \ +2.7 \times 10^{-16} \exp{(-h/1500)}+A \exp{(-h/100)},
\end{aligned}
\end{equation}
where $h$ denotes the altitude, $v_{\rm{rms}}$ is the rms wind speed in m/s, and $A=C^2_n (0)$. In this work, we consider the Kolmogorov model~\cite{K41} for the atmospheric turbulence.
\begin{figure}[ht!]
	\centering
	\includegraphics[scale=0.3]{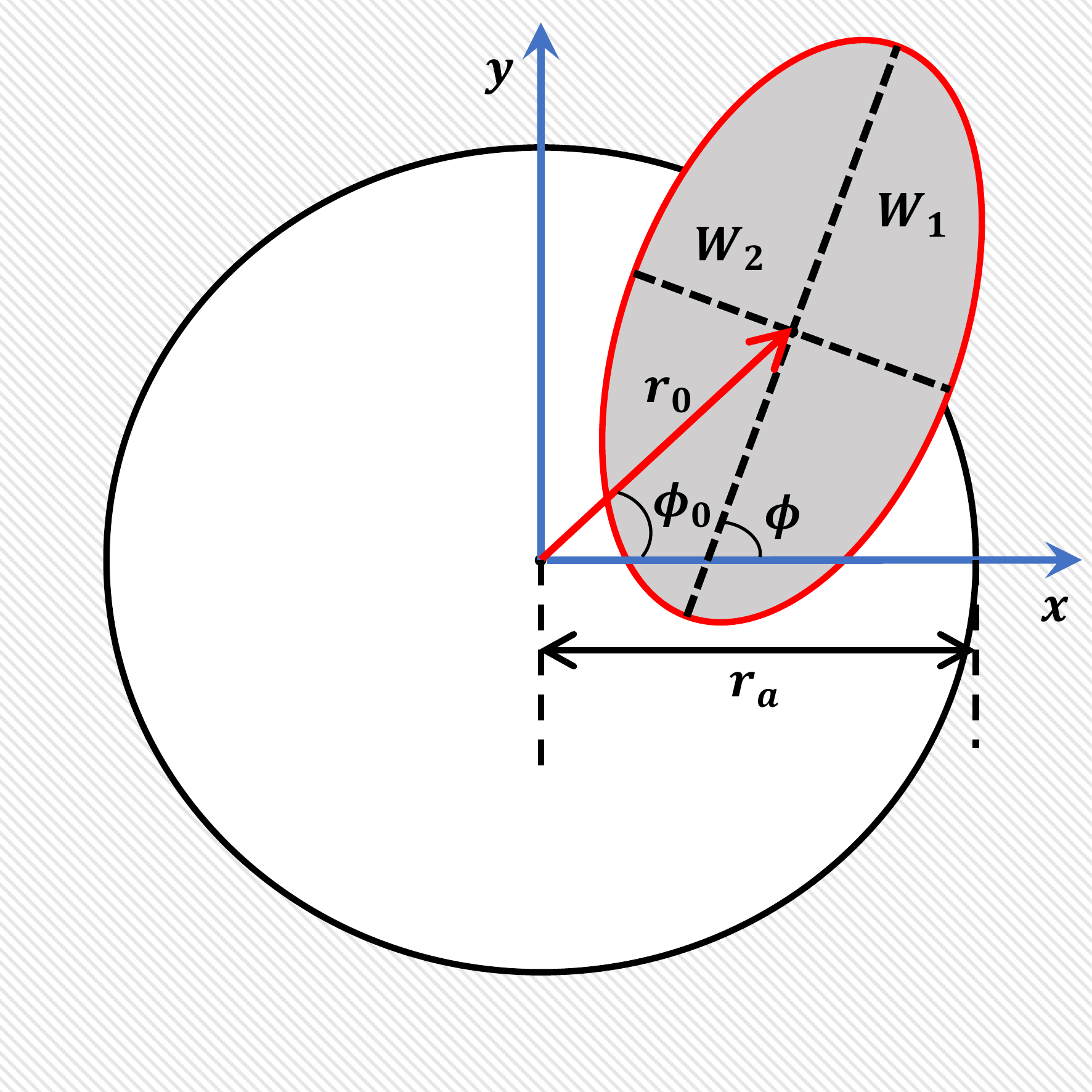}
	\caption{Illustration of the atmospheric turbulence effect on the intensity profile. The white area represents the receiver aperture (with an aperture radius of $r_a$), and the shaded area represents the turbulence-affected intensity profile of a beam. In this figure, $r_0$ and $\phi_0$ characterize the deviation of the wandering beam centroid. $W_1$ and $W_2$ are elliptical semi-axes lengths, and $\phi$ denotes the rotation angle of the beam.}
	\label{fig:beam_wandering}
\end{figure}

\begin{figure}[ht!]
	\centering
	\vspace{-6pt}
	\hspace{8pt}
	\includegraphics[scale=0.04]{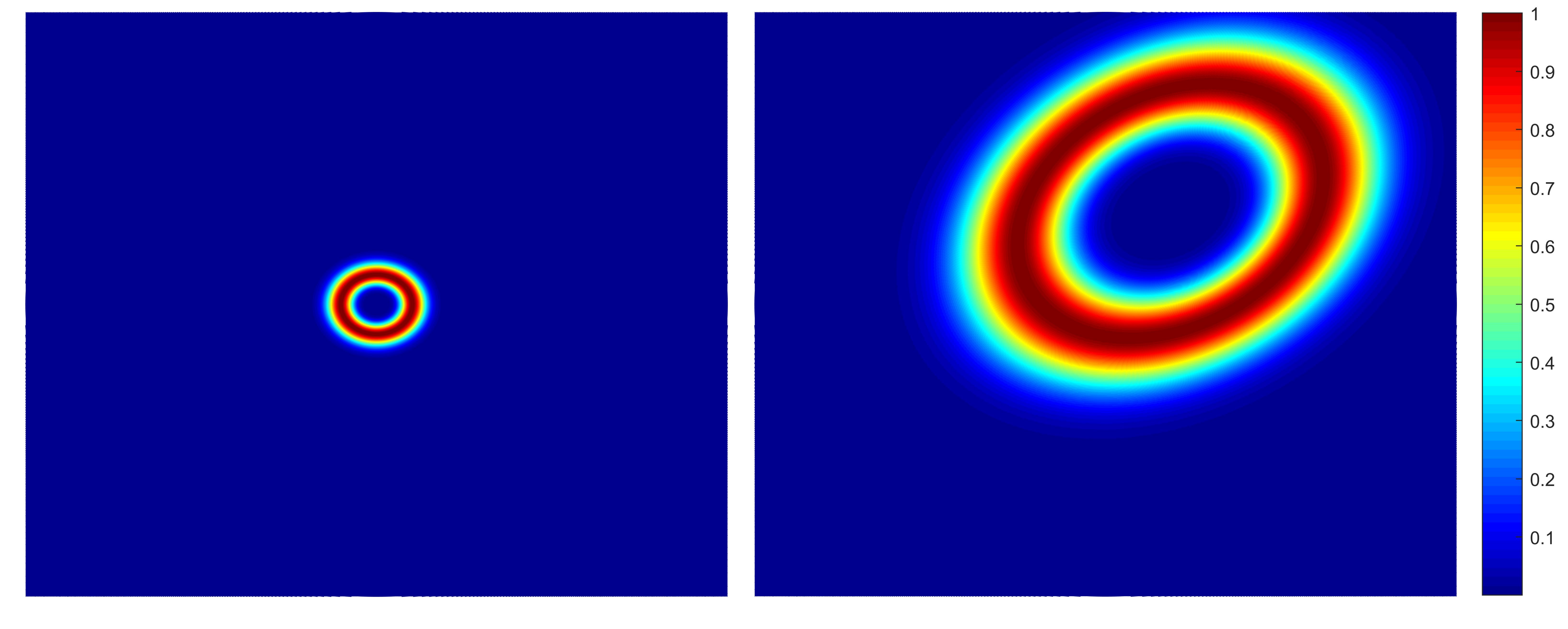}
	\vspace{-5pt}
	\caption{The normalized intensity profile of the $LG_{04}$ mode before (left), and after (right) the atmospheric turbulence effect.}
	\label{Patterns_BWDF}
\end{figure}

The atmospheric turbulence leads to three effects on the intensity profile, namely beam wandering, beam broadening, and elliptical-shape beam deformation (as illustrated in Fig.~(\ref{fig:beam_wandering})). These three effects were well described in~\cite{beam_wandering2016} where good agreements were found with the experimental results in~\cite{PDT1.6km_1,PDT1.6km_2,143km}. The three effects mentioned above are characterized by five random parameters, $\{r_0, \phi_0, W_{1}, W_{2}, \phi\}$. The parameters $\{r_0,\phi_0\}$ characterize the centroid wandering of the beam, and $\{W_{1},W_{2},\phi\}$ characterize the elliptical-shape deformation and the broadening of the beam (see Fig.~(\ref{fig:beam_wandering})). The distributions of these five parameters have been well established for non-OAM-beams using the Gaussian approximation under the assumption that the atmospheric turbulence is isotropic~\cite{beam_wandering2016}. Specifically, $\phi$ is uniformly distributed over $[0,\pi/2)$, and is independent of the other parameters. The wandering beam-centroid position (in Cartesian coordinates), $x_0 = r_0\cos{\phi_0}$ and $y_0 = r_0\sin{\phi_0}$, can be considered as independent zero-mean Gaussian variables. Finally, $W_{1}$ and $W_{2}$ follow a multi-variable log-normal distribution. Denoting $\Theta_1=\ln{\frac{W_{1}^2}{w_0^2}}$, and $\Theta_2=\ln{\frac{W_{2}^2}{w_0^2}}$, $\{x_0, y_0,\Theta_1,\Theta_2\}$ follow a joint Gaussian distribution with mean values of
\begin{equation}\label{Eq.DistributionMean_BW}
\begin{gathered}
\left\langle x_0\right\rangle =\left\langle y_0\right\rangle= 0, \\
\left\langle\Theta_{1}\right\rangle =\left\langle\Theta_{2}\right\rangle =\ln \left[\frac{\left(1+2.96 \sigma_{I}^{2} \Omega^{5 / 6}\right)^{2}}{\Omega^{2} \sqrt{\left(1+2.96 \sigma_{I}^{2} \Omega^{5 / 6}\right)^{2}+1.2 \sigma_{I}^{2} \Omega^{5 / 6}}}\right],
\end{gathered}
\end{equation}
and a covariance matrix of
\begin{equation}\label{Eq.CovM}
\mathcal{M}=\left[ \begin{array}{cccc}{\left\langle x_0 ^{2}\right\rangle} & { 0} & {0} & {0} \\ {0} & {\left\langle y_0^{2}\right\rangle} & { 0} & {0} \\ {0} & {0} & {\left\langle\Theta_{1}^{2}\right\rangle} & {\left\langle\Theta_{1} \Theta_{2}\right\rangle} \\ { 0} & {0} & {\left\langle\Theta_{1} \Theta_{2}\right\rangle} & {\left\langle\Theta_{2}^{2}\right\rangle}\end{array}\right].
\end{equation}
The elements in Eq.~(\ref{Eq.DistributionMean_BW}) and Eq.~(\ref{Eq.CovM}) are given by
\begin{equation}\label{Eq.CovM_Elements}
\begin{aligned}
\left\langle x_{0}^{2}\right\rangle =\left\langle y_{0}^{2}\right\rangle&= 0.33 w_{0}^{2} \sigma_{I}^{2} \Omega^{-7 / 6}, \\
\left\langle\Theta_{1}^{2}\right\rangle =\left\langle\Theta_{2}^{2}\right\rangle &=\ln \left[1+\frac{1.2 \sigma_{I}^{2} \Omega^{5 / 6}}{\left(1+2.96 \sigma_{I}^{2} \Omega^{5 / 6}\right)^{2}}\right],
\\\left\langle\Theta_{1} \Theta_{2}\right\rangle \ \ \ &=\ln \left[1-\frac{0.8 \sigma_{I}^{2} \Omega^{5 / 6}}{\left(1+2.96 \sigma_{I}^{2} \Omega^{5 / 6}\right)^{2}}\right],
 \end{aligned}
\end{equation}
where $\Omega={kw_{0}^{2}}/{2L}$ with $L$ being the propagation distance, and $\sigma_I^2$ is the scintillation index. For satellite-to-ground (downlink) channels, $\sigma_I^2$ can be calculated by~\cite{book}
\begin{equation}\label{Eq.ScintIdx}
\sigma_{I}^{2}=\exp \left[\frac{0.49 \sigma_{R}^{2}}{\left(1+1.11 \sigma_{R}^{12 / 5}\right)^{7 / 6}}+\frac{0.51 \sigma_{R}^{2}}{\left(1+0.69 \sigma_{R}^{12 / 5}\right)^{5 / 6}}\right]-1,
\end{equation}
with $\sigma_{R}^2$ being the Rytov variance,
\begin{equation}\label{RytovVar}
\sigma_{R}^{2}=2.25 k^{7 / 6} \sec ^{11 / 6}(\theta_z) \int_{h_{0}}^{H} C_{n}^{2}(h)\left(h-h_{0}\right)^{5 / 6} d h.
\end{equation}
A before-and-after comparison of the atmospheric turbulence effect on the intensity profile of the $LG_{04}$ mode is shown in Fig.~(\ref{Patterns_BWDF}) for illustration. Note that for the $LG_{0l}$ mode, $W_1$ and $W_2$ need to be scaled by a factor of $\sqrt{|l|+1}$ to faithfully characterize the beam divergence.

The atmospheric turbulence effect on the phase profile can be modeled as a single random phase screen $\alpha (r,\theta)$ in the receiver aperture plane~\cite{PatersonSPS}. The phase screen $\alpha (r,\theta)$ manifests itself as a phase perturbation term of $\exp{[i\alpha (r,\theta)]}$ in the complex amplitude of the modified beam.  The function $\alpha (r,\theta)$ can be statistically characterized by its phase structure function
\begin{equation}\label{Eq.PSF}
\mathcal{D}_{\alpha}(r,\theta,r^{\prime},\theta^{\prime})=\left\langle|\alpha(r,\theta)-\alpha(r^{\prime},\theta^{\prime})|^{2}\right\rangle,
\end{equation}
where $(r,\theta)$ and $(r^{\prime},\theta^{\prime})$ denote two points in the aperture plane. $\mathcal{D}_{\alpha}(r,\theta,r^{\prime},\theta^{\prime})$ can be expressed in a simplified form of
\begin{equation}\label{Eq.PSF_Fried}
\mathcal{D}_{\alpha}(r,\theta,r^{\prime},\theta^{\prime})=D_{\alpha}(|\Delta r|)=6.88\left(\frac{|\Delta r|} {r_{F}}\right)^{5 / 3},
\end{equation}
where $\Delta r$ is the separation distance between $(r,\theta)$ and $(r^{\prime},\theta^{\prime})$, and $r_{F}$ is the well-known Fried parameter~\cite{Fried66} that characterizes the phase coherence of the turbulent atmosphere. This parameter is given by~\cite{book_near_earth_laser}
\begin{equation}\label{Eq.FriedParam}
r_{F}=\left[0.423 k^{2} \sec \theta_z \int_{h_0}^{H} C_{n}^{2}(h) d h\right]^{-3 / 5},
\end{equation}
for satellite-to-ground channels.
\subsection{Evolution of the perturbed state}\label{BeamEvo}
Considering the atmospheric turbulence effect introduced in Section~\ref{channel_model}, we now express $\Psi(r, \theta, z)$ as
\begin{equation}\label{Eq.Rx}
\Psi(r, \theta, z)=H(r) \Psi_D(r, \theta, z) \exp [i \alpha (r, \theta)].
\end{equation}
In the above equation, $H(r)$ is the symmetrical circular aperture function which is given by
\begin{equation}\label{Eq.Aperature}
H(r)=\left \{
\begin{array}
{l}{1, \ \ \ r \leq r_{a}} \\
{0, \ \ \ r>r_{a}}\end{array}, \right.
\end{equation}
and $\Psi_D(r, \theta, z)$ is given by
\begin{equation}\label{Eq.Perturbed_D}
\Psi_D ( r , \theta , z ) = \frac{\varphi_{0,l_0} ( r_i , \theta_i, z )}{\sqrt{|\mathbf{S}|}},
\end{equation}
where
\begin{equation}\label{Eq.R'}
\begin{gathered}
\begin{aligned}
r_i=&\sqrt{(x_i)^{2}+(y_i)^{2}}, \\ \theta_i=&\atantwo(y_i,x_i).
\end{aligned}
\end{gathered}
\end{equation}
The parameters in Eq.~(\ref{Eq.Perturbed_D}) and Eq.~(\ref{Eq.R'}) are given by
\begin{equation}\label{}
\left[ \begin{array}{l}{x_i} \\ {y_i}\end{array}\right]=S^{-1} \left[ \begin{array}{l}{x_{t}} \\ {y_{t}}\end{array}\right],
\end{equation}
\begin{equation}
\begin{aligned}
x_t &= r \cos\theta - r_0\cos\phi_0, \\
y_t &= r \sin\theta - r_0\sin\phi_0,
\end{aligned}
\end{equation}
where $\mathbf{S}$ is a matrix that characterizes the beam shape, and it is given by
\begin{equation}
\mathbf{S}=\frac{1}{r_{0 l_0}(z)} \left[\begin{matrix}{W_{1} \cos \phi} & {-W_{2} \sin \phi} \\ {W_{1} \sin \phi} & {\ \ W_{2} \cos \phi}\end{matrix}\right].
\end{equation}

\subsection{Evaluation of OAM detection probability}
After expressing $\Psi(r, \theta, z)$ as Eq.~(\ref{Eq.Rx}), we are now in a position to evaluate $P(l_0)$ and $P(l_r)$. First, we substitute Eq.~(\ref{Eq.Rx}) into Eq.~(\ref{Eq.RFC}), and rewrite the RFC function as
\begin{equation}\label{Eq.RFC_Final}
\begin{aligned}
C_{\Psi}(r, \theta,\theta^{\prime},z)&=\left\langle \Psi_D ^ { * } ( r ,\theta^{\prime}, z ) \Psi_D ( r , \theta , z ) \right\rangle \\
&\times C_\alpha (r,\theta,\theta^{\prime}) \times H^2(r),
\end{aligned}
\end{equation}
where $C_\alpha (r,\theta,\theta^{\prime})$ is given by
\begin{equation}\label{Eq.PhaseCorr}
C_\alpha (r,\theta,\theta^{\prime}) =\langle\exp \{i[\alpha(r, \theta)-\alpha(r, \theta^{\prime})]\}\rangle.
\end{equation}
Assuming the refractive index fluctuations to be a Gaussian random process so that $\langle\exp(ix)\rangle=\exp\left(-\frac{1}{2}\left\langle|x|^{2}\right\rangle\right)$ is true~\cite{PatersonSPS}, we can express $C_\alpha (r,\theta,\theta^{\prime})$ as
\begin{equation}\label{Eq.Phase_corr_final}
\begin{aligned}
C_\alpha (r,\theta,\theta^{\prime})=&\exp \!\left[-\frac{1}{2} D_{\alpha}\left(\left|2 r \sin \left(\frac{\theta-\theta^{\prime}}{2}\right)\right|\right)\right] \\
=&\exp \!\left[-6.88 \! \times \! 2^{2 / 3} \! \left(\!\frac{r}{r_{F}}\!\right)^{5 / 3}\!\left|\sin \left(\frac{\theta-\theta^{\prime}}{2}\right)\!\right|^{5 / 3}\right]\!\!.
\end{aligned}
\end{equation}
Substituting Eqs.~(\ref{Eq.Aperature}),~(\ref{Eq.Perturbed_D}),~(\ref{Eq.Phase_corr_final}) into Eq.~(\ref{Eq.RFC_Final}), and then substituting Eq.~(\ref{Eq.RFC_Final}) into Eq.~(\ref{Eq.ProbFinal}), we can evaluate $P(l_0)$ and $P(l_r)$.

In this work, we carry out Monte Carlo simulations to evaluate $P(l_0)$ and $P(l_r)$. First, we generate a large number of realizations of the atmospheric turbulence by randomly generating their characterizing parameters $\{r_0, \phi_0, W_{1}, W_{2}, \phi\}$ following the distributions discussed in Section~\ref{channel_model}. Then, for each realization we numerically evaluate $\Psi_D ( r , \theta , z )$ and $C_\alpha (r,\theta,\theta^{\prime})$ to give $C_{\Psi} (r,\theta,\theta^{\prime},z)$. Note that while evaluating $C_\alpha (r,\theta,\theta^{\prime})$, $r_F$ is fixed, since $r_F$ is not dependent on specific realizations of the atmospheric turbulence (see Eq.~(\ref{Eq.FriedParam})). Afterwards, realizations of $C_{\Psi} (r,\theta,\theta^{\prime},z)$ are used to obtain realizations of $P(l_0)$ and $P(l_r)$. At last, all realizations of $P(l_0)$ and $P(l_r)$ are averaged to give their final values.

\section{Simulations}\label{ch:Sim}
\subsection{Simulation settings}
In our simulation, we generate 2000 independent realizations of the atmospheric turbulence, and then evaluate the corresponding OAM detection performances. The final result of $P(l_0)$ and $P(l_r)$ is acquired by averaging over all the realizations.

For the system settings, we consider a vertical satellite-to-ground channel (i.e. $\theta_z=0$), and thus we have $H=h_0+L$. We restrict our study to the Low Earth Orbit (LEO) setting, and set the maximum satellite altitude to $H_{\rm{max}}=500\rm{km}$. We consider three cases of $h_0=0\rm{m}$, $h_0=1000\rm{m}$, and $h_0=3000\rm{m}$, giving maximum channel distances of $L_{\rm{max}}=500\rm{km}$, $L_{\rm{max}}=499\rm{km}$, and $L_{\rm{max}}=497\rm{km}$, respectively. For the atmospheric channel parameters, we set $C_{n}^{2}(0)=9.6 \times 10^{-14} \mathrm{m}^{-2 / 3}$, and $v_{\rm{rms}}=6\rm{m/s}$. These atmospheric channel parameters accord with a realistic setting which was adopted in~\cite{Submarine_CVQKD} to study satellite-based QKD.

For the optical parameters, we are interested in two optical wavelengths, $\lambda=1550 \rm{nm}$ and $\lambda=800 \rm{nm}$, since they are currently widely used in FSO communications. For all the transmitted $LG_{0l}$ modes, $w_0$ is fixed to be $15{\rm{cm}}$. We set $r_{t} = r_{0l_{\rm{max}}}(0)$, and $r_{a} = r_{0l_{\rm{max}}}(L_{\rm{max}})$ in order to achieve reasonable transmission and detection efficiencies. In these simulations, we consider a Hilbert space of $l_{\rm{max}}=4$, leading to $r_{t}=33$cm and $r_{a}=3.7$m (the aperture radius that achieves effectively zero loss due to diffraction). The values of $r_{t}$ and $r_{a}$ chosen in our simulation are reasonable in practice.

Since the spatial phase structure $\exp(il\theta)$ plays an important role in the OAM detection, it is intuitive to think that the adaptive optics (AO) techniques (in the following we will assume phase-only AO techniques) could be helpful in improving the detection performance. As such, in our simulations, we also consider the performance improvement provided an AO system. The AO system with actuator deformable mirrors and Shack-Hartmann wavefront sensors has been used previously (e.g. in~\cite{OAM_AO1,Rodenburg14,OAM_AO2}) to improve the OAM detection performance. Specifically,~\cite{OAM_AO2} sampled the atmospheric turbulence using a beacon beam with a different wavelength, and showed that the wavefront phase distortions of $LG_{0l}$ modes can be significantly compensated (in a real-time fashion) using a properly designed closed-loop AO system.

\subsection{Simulation results}\label{ch:SimResults}
Before presenting our simulation results it should be pointed out that in all our results we observe that the OAM detection performance is dependent on $|l_0|$. An OAM eigenstate with a larger $|l_0|$ suffers from a lower detection probability, and a more severe crosstalk. Such observations accord with realistic experimental results in the literature (e.g.~\cite{Ren_13,Rodenburg14}), indicating that the atmospheric channel model considered in this work is realistic. We also observe that the detection performances for OAM eigenstates with $\pm l_0$ are almost equal, since these eigenstates differ only in the handedness of their azimuthal  structures (see~Eq.~(\ref{Eq.OAM_Eigenstate})).

\begin{figure}[ht!]
	\centering
	\includegraphics[scale=0.6]{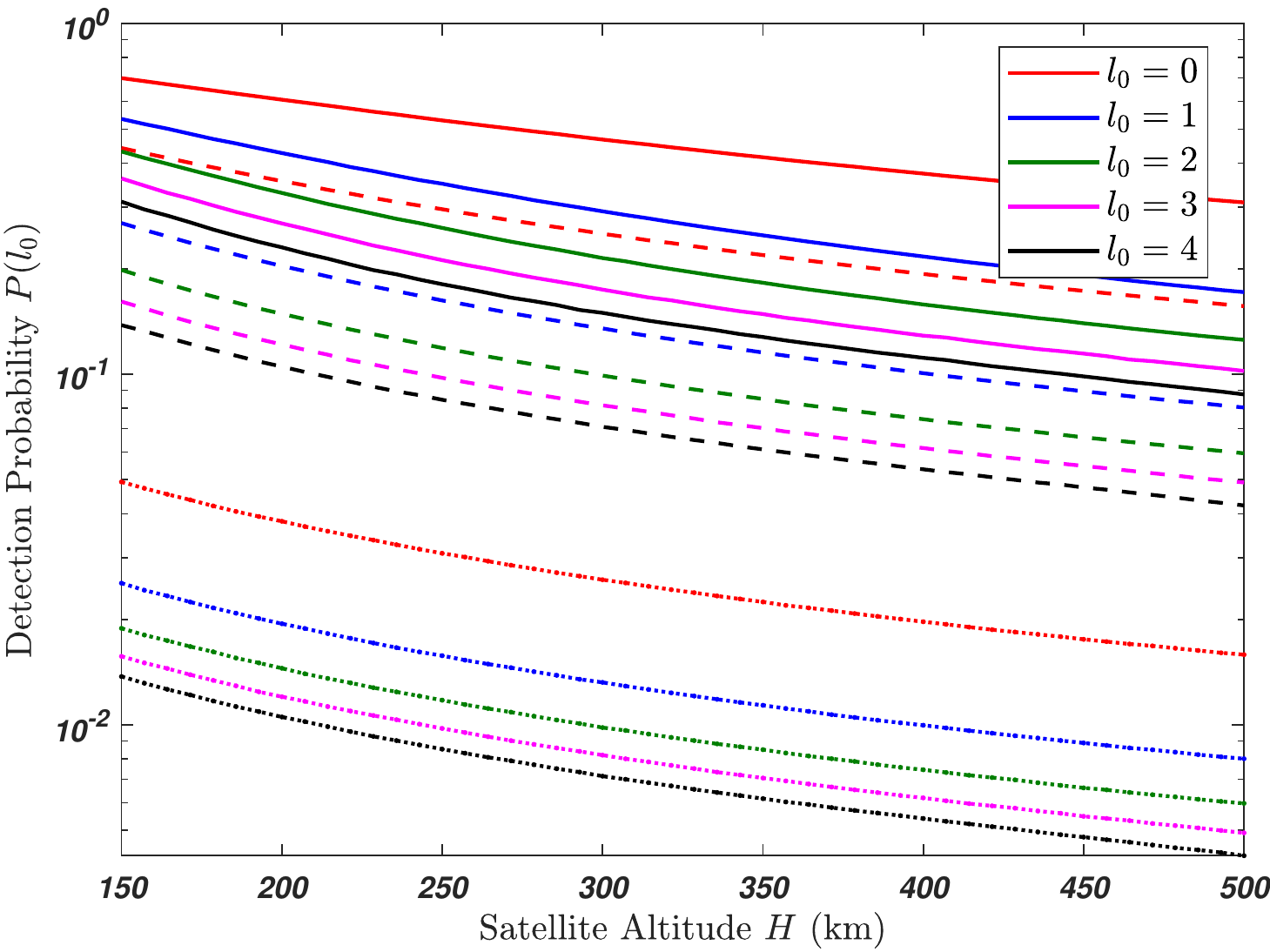}
	\caption{Detection probabilities under $h_0=0$m (dash-dot), $h_0=1000$m (dashed), and $h_0=3000$m (solid). Here we set $\lambda=1550$nm.}
	\label{fig:alt1550}
\end{figure}

\begin{figure}[ht!]
	\centering
	\includegraphics[scale=0.6]{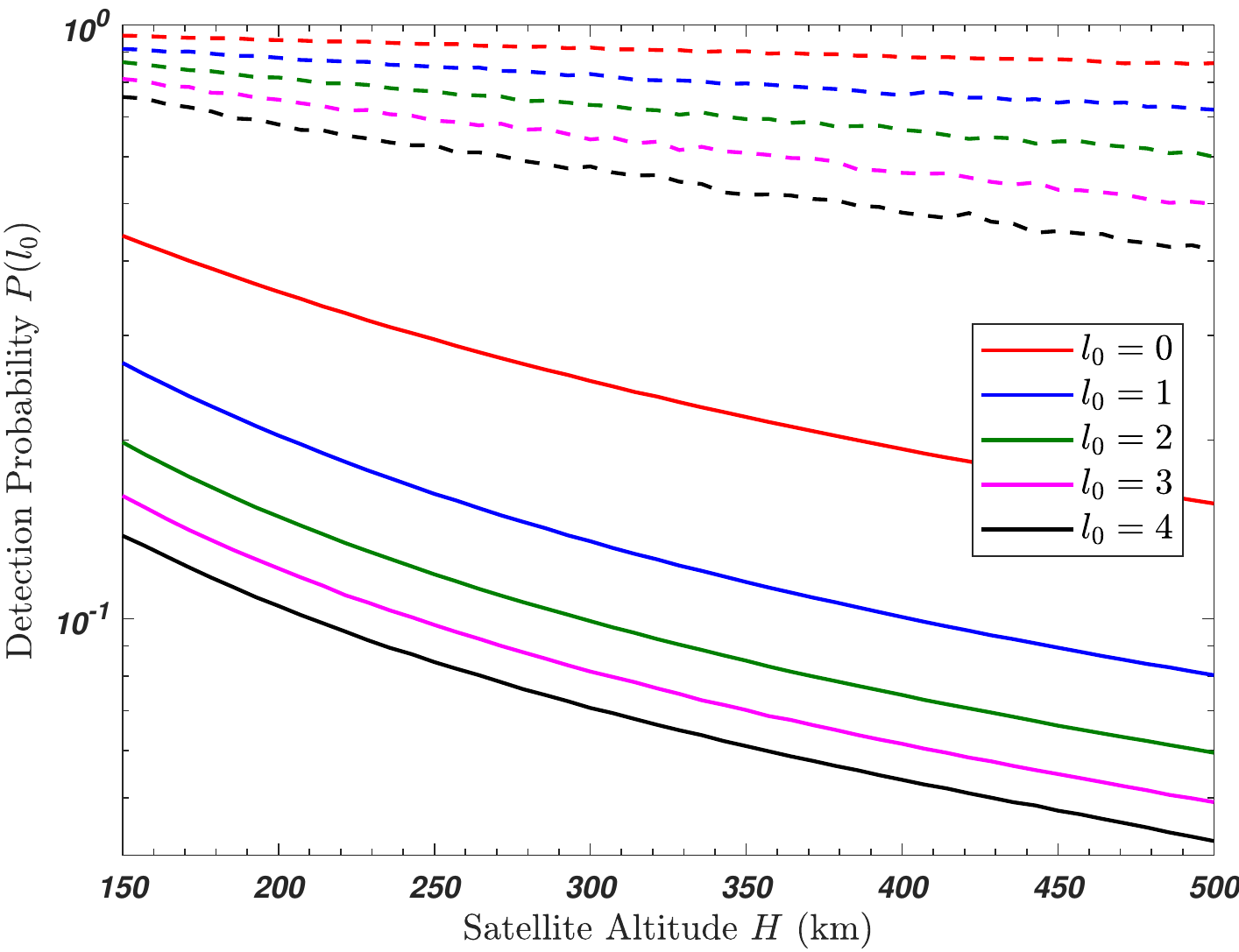}
	\caption{Detection probabilities achieved with (dashed), and without (solid) AO techniques. Here we set $h_0=1000$m, and $\lambda=1550$nm.}
	\label{fig:AO}
\end{figure}

\begin{figure}[ht!]
	\centering
	\includegraphics[scale=0.6]{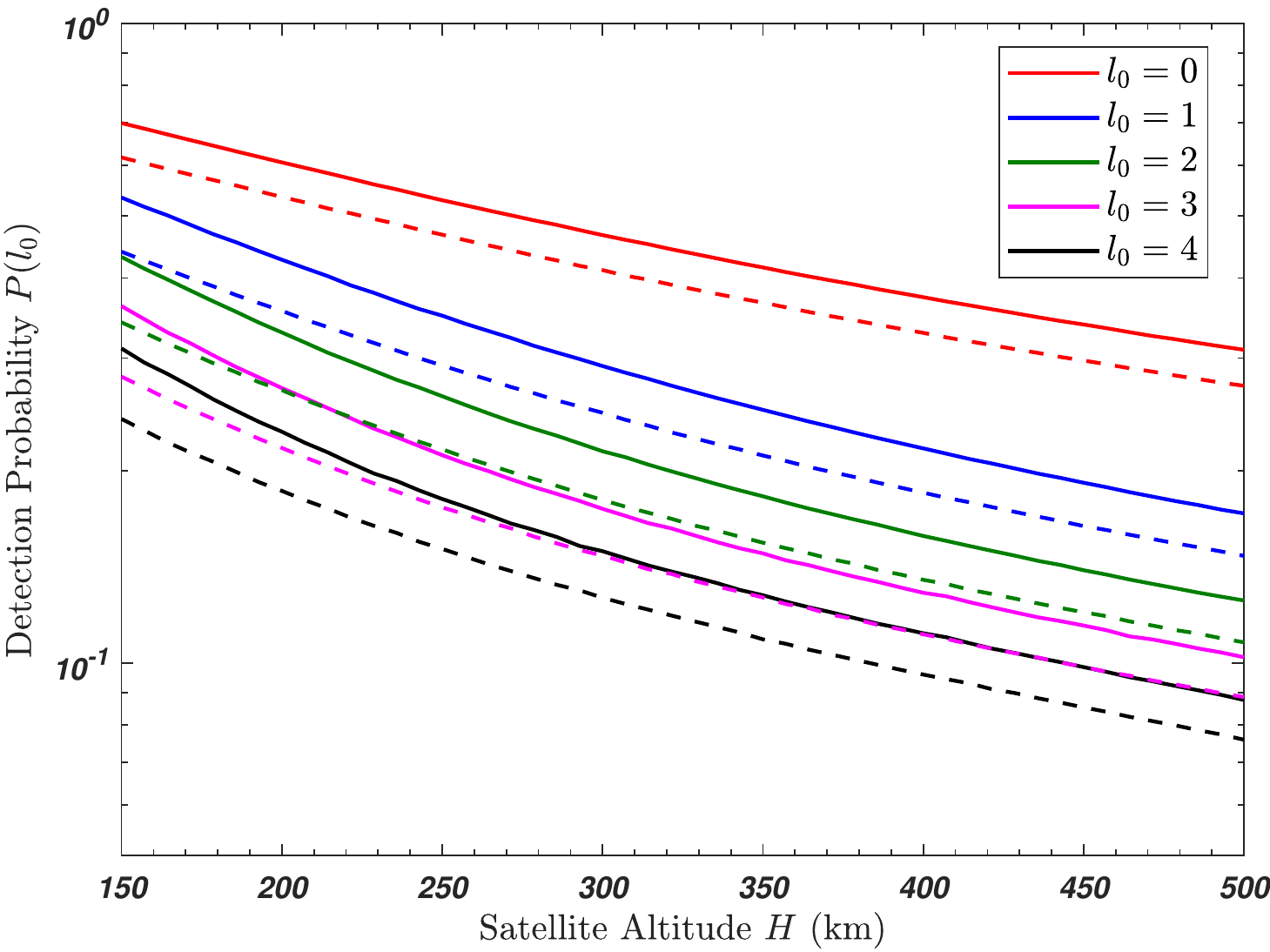}
	\caption{Detection probabilities at $\lambda=800$nm (dashed), and $\lambda=1550$nm (solid). Here we set $h_0=3000$m. No AO techniques are considered.}
	\label{fig:wavelength}
\end{figure}

First, we consider an OAM-based system utilizing a Hilbert space of $l_{\rm{max}}=4$, and present in Figs.~(\ref{fig:alt1550})~--~(\ref{fig:wavelength}) the simulation results for the OAM detection probability as a function of the satellite altitude. In these figures, for a given $h_0$ and $\lambda$, the detection probability decreases as the satellite altitude increases, since the atmospheric turbulence effect becomes more severe.

In Fig.~(\ref{fig:alt1550}), we show the detection probabilities for OAM eigenstates with different $l_0$ values under  different $h_0$ values. In this figure, we can see that a ground station placing at a higher altitude is more preferable for satellite-to-ground OAM quantum communications. This preference is due to the fact that the turbulence strength decreases rapidly with altitude. However, even under $h_0=3000 \rm{m}$, at $H=500 \rm{km}$  we observe poor detection probabilities of $P(l_0)=0.31$ for $l_0=0$, $P(l_0)=0.17$ for $l_0=1$, $P(l_0)=0.13$ for $l_0=2$, $P(l_0)=0.10$ for $l_0=3$, and $P(l_0)=0.09$ for $l_0=4$. Such detection probabilities may only allow for limited discrimination of the received OAM states.

In Fig.~(\ref{fig:AO}), we compare the detection probabilities achieved with and without phase-only AO techniques. Here we assume an ideal AO system that can perfectly correct the atmospheric turbulence effect on the phase profile. The results shown in this figure accord with the results presented in~\cite{OAM_AO1,OAM_AO2,Rodenburg14}, confirming the potential to use AO techniques in OAM  satellite-to-ground quantum communications for a considerable performance boost. Specifically, under $h_0=1000 \rm{m}$, at $H=500 \rm{km}$ we still observe good detection probabilities of $P(l_0)=0.87$ for $l_0=0$, $P(l_0)=0.72$ for $l_0=1$, $P(l_0)=0.60$ for $l_0=2$, $P(l_0)=0.50$ for $l_0=3$, and $P(l_0)=0.42$ for $l_0=4$ with the help of pure-phase AO techniques. Such detection probabilities are sufficient to achieve reasonably good
discrimination of the received OAM states.

From Fig.~(\ref{fig:AO}), we can also see that the correction of errors on the phase profile plays a  significant role, as indicated by the dashed curves based on the use of AO techniques (the degradation in the OAM detection probability caused by the  turbulence  on the intensity profile is considerably smaller). Indeed, in satellite-to-ground atmospheric channels, the atmospheric turbulence layer is only in the vicinity of the receiver, making beam wandering and elliptical-shape deformation effects rather insignificant.\footnote{Following transmission from the satellite, the beam size at entry into the turbulence layer  is generally larger than the scale of the turbulent eddies~\cite{book}. Also, the beam-broadening effect is mainly due to pure diffraction in satellite-to-ground channels~\cite{book}.} Therefore, the main contributors to the degradation of OAM detection performance in satellite-to-ground channels are the diffraction-induced photonic losses and the atmospheric turbulence effect on the phase profile. In our simulations, the diffraction-only induced photonic losses can be neglected due to our assumption of a large receiver aperture.

In Fig.~(\ref{fig:wavelength}), we compare the OAM detection probabilities achieved at different optical wavelengths. We can see that a larger wavelength is preferable. This is due to the fact that a larger wavelength reduces the atmospheric turbulence effect on both the intensity profile and the phase profile (one could easily show that $\sigma_{R}^{2} \propto \lambda^{-7/6}$ and $r_{0l_0}/ r_F \propto \lambda^{-1/5}$).

\begin{figure}[th!]
	\centering
	\includegraphics[scale=0.09]{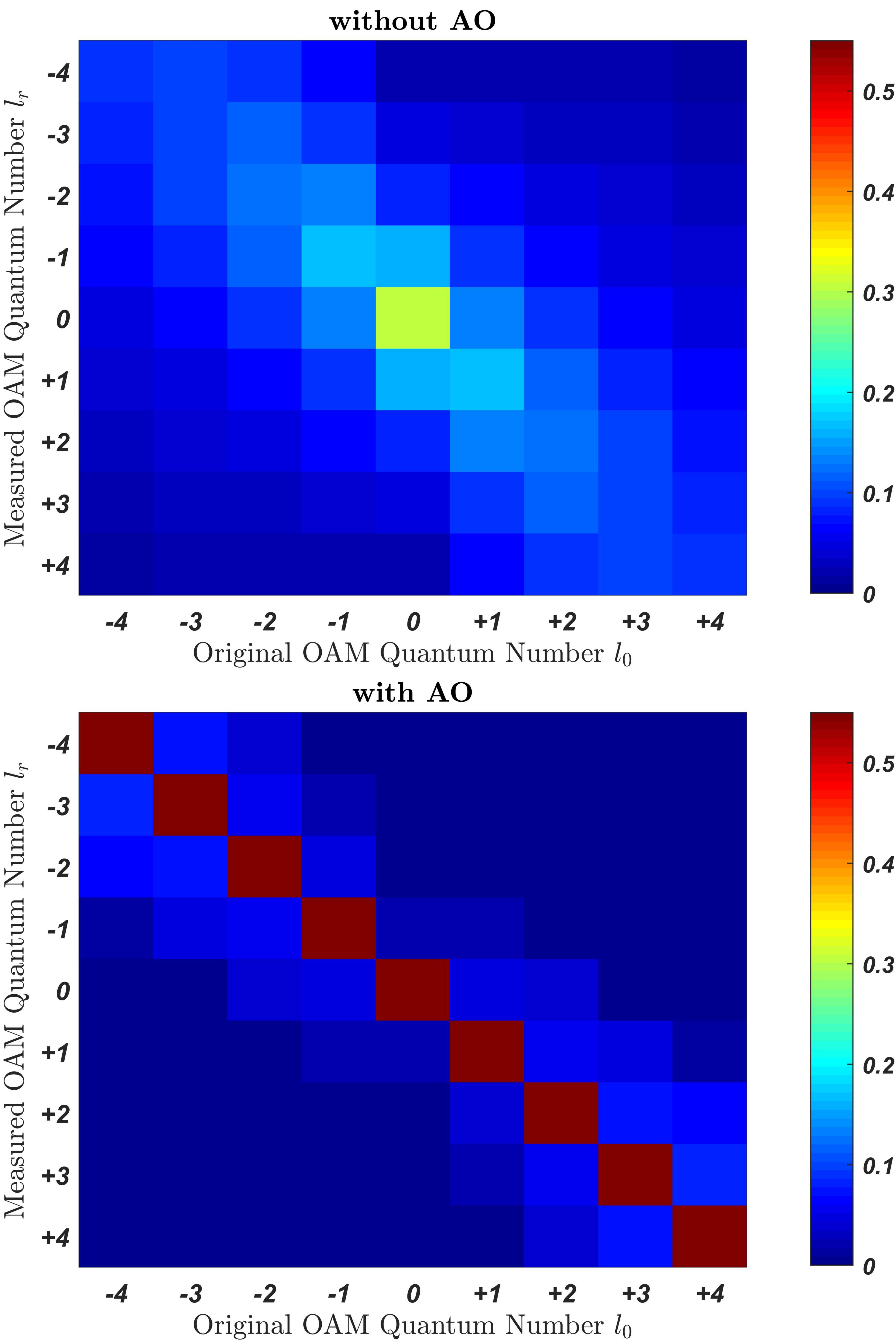}
	\caption{Crosstalk matrices of two systems without (top) and with (bottom) AO techniques. Here we set $H=500 \rm{km}$, $h_0=3000m$, and $\lambda=1550$nm.}
	\label{fig:Corsstalk}
\end{figure}

Now we consider two systems, with and without AO techniques, utilizing a Hilbert space of $l_{\rm{max}}=4$, and show in Fig.~(\ref{fig:Corsstalk}) their corresponding channel crosstalk matrices under $h_0=3000\rm{m}$ and $H=500 \rm{km}$. From both sub-figures, we observe that OAM eigenstates with $\pm l_0$ show very similar crosstalk behaviors. We also observe that the atmospheric turbulence preferentially introduces similar degrees of crosstalk into neighboring OAM eigenstates (e.g. $P{(l_r=1)}\simeq P{(l_r=-1)}$ for $l_0=0$, and $P{(l_r=2)}\simeq P{(l_r=0)}$ for $l_0=1$), giving the crosstalk matrices a symmetrical look. After adopting ideal pure-phase AO techniques, we can see the detection probabilities are considerably enhanced. Specifically we have $P(l_0)=0.94$ for $l_0=0$, $P(l_0)=0.87$ for $l_0=1$, $P(l_0)=0.79$ for $l_0=2$, $P(l_0)=0.72$ for $l_0=3$, and $P(l_0)=0.65$ for $l_0=4$, indicating that effective discrimination of OAM states could be achieved. Also, we can see that the crosstalk is significantly reduced after adopting AO techniques.

\section{Conclusions}
The atmospheric turbulence within the channel imposes undesirable decoherence effects on OAM-beams, and limits the practical usage of OAM. The detection performance of OAM states determines the feasibility of OAM in satellite-based quantum communications, since it directly links to the quantum bit error rate, and the evolution of OAM entanglement.
In this work, we validated the feasibility of OAM in satellite-to-ground quantum communications by determining the OAM detection performance using Monte Carlo simulations. Specifically, we determined the parameters under which the use of OAM is feasible for satellite-to-ground quantum communications. In carrying out our calculations, we determined a realistic model for the satellite-to-ground atmospheric channel considering the atmospheric turbulence effect on both the intensity and phase profiles of the OAM-beams. We showed that, although the atmospheric turbulence has undesirable effects on OAM detection, these effects could be alleviated by placing the ground station at a higher altitude, choosing a suitable wavelength, and\slash or adopting advanced real-time AO techniques. Our work provides new insights into future implementations of space-based OAM systems within the context of quantum communications.

\bibliographystyle{IEEEtran}

\end{document}